\documentclass[aps,pra,twocolumn,nopacs,lettersize,superscriptaddress]{revtex4-1}
\usepackage{array}
\usepackage{rotating}
\usepackage{multirow}
\usepackage{placeins}
\usepackage{xcolor}
\usepackage{hyperref}
\usepackage{soul}
\hypersetup{
    colorlinks=true,
    linkcolor=blue,
    citecolor=blue,
    citebordercolor=white,
    linkbordercolor=white,
}
\usepackage[margin=0.65in]{geometry}
\usepackage{graphicx}
\graphicspath{ {} }
\usepackage{float}
\usepackage{amsmath,amssymb,amsfonts}
\usepackage{natbib}

\begin{document}

\title{Giant Vortex Clusters in a Two-Dimensional Quantum Fluid}

\author{Guillaume Gauthier}
\thanks{These authors contributed equally to this work.}
\affiliation{Australian Research Council Centre of Excellence for Engineered Quantum Systems, School of Mathematics and Physics, University of Queensland, St. Lucia, QLD 4072, Australia.}
\author{Matthew T. Reeves}
\thanks{These authors contributed equally to this work.}
\affiliation{Australian Research Council Centre of Excellence in Future Low-Energy Electronics Technologies, School of Mathematics and Physics, University of Queensland, St Lucia, QLD 4072, Australia.}
\author{Xiaoquan Yu}
\affiliation{Department of Physics, Centre for Quantum Science, and Dodd-Walls Centre for Photonic and Quantum Technologies, University of Otago, Dunedin, New Zealand.}
\author{Ashton S. Bradley}
\affiliation{Department of Physics, Centre for Quantum Science, and Dodd-Walls Centre for Photonic and Quantum Technologies, University of Otago, Dunedin, New Zealand.}
\author{Mark Baker}
\affiliation{Australian Research Council Centre of Excellence for Engineered Quantum Systems, School of Mathematics and Physics, University of Queensland, St. Lucia, QLD 4072, Australia.}
\author{Thomas A. Bell}
\affiliation{Australian Research Council Centre of Excellence for Engineered Quantum Systems, School of Mathematics and Physics, University of Queensland, St. Lucia, QLD 4072, Australia.}
\author{Halina Rubinsztein-Dunlop}
\affiliation{Australian Research Council Centre of Excellence for Engineered Quantum Systems, School of Mathematics and Physics, University of Queensland, St. Lucia, QLD 4072, Australia.}
\author{Matthew J. Davis}
\affiliation{Australian Research Council Centre of Excellence in Future Low-Energy Electronics Technologies, School of Mathematics and Physics, University of Queensland, St Lucia, QLD 4072, Australia.}
\author{Tyler W. Neely}
\affiliation{Australian Research Council Centre of Excellence for Engineered Quantum Systems, School of Mathematics and Physics, University of Queensland, St. Lucia, QLD 4072, Australia.}

\date{\today}

\begin{abstract}
\noindent Adding energy to a system through transient stirring usually leads to more disorder. In contrast, point-like vortices in a bounded two-dimensional fluid are predicted to reorder above a certain energy, forming persistent vortex clusters. Here we realize experimentally these vortex clusters in a planar superfluid: a $^{87}$Rb Bose-Einstein condensate confined to an elliptical geometry. We demonstrate that the clusters persist for long times, maintaining the superfluid system in a high energy state far from global equilibrium. Our experiments explore a regime of vortex matter at negative absolute temperatures, and have relevance to the dynamics of topological defects, two-dimensional turbulence, and systems such as helium films, nonlinear optical materials, fermion superfluids, and quark-gluon plasmas.
\end{abstract}
\maketitle

An isolated system that is initially stirred will in most cases eventually achieve quiescent thermodynamic equilibrium. However, in some systems, the near decoupling of particular degrees of freedom can result in an isolated subsystem with a different time-scale for equilibration~\cite{kraichnan1980two}. Strikingly, the subsystem can exhibit highly-correlated and non-uniform thermal equilibria ~\cite{Montgomery:1974dra,kraichnan1967inertial,onsager1949statistical}.   As recognized by Lars Onsager~\cite{onsager1949statistical} a prototypical example is a system of N point vortices~\cite{lin1941motion} contained within a bounded two-dimensional (2D) fluid. This model predicts that, given sufficient decoupling between two and three-dimensional flow and negligible viscous dissipation, high energy fluid flow yields low-entropy equilibria that exhibit large-scale aggregations of like-circulation vortices~\cite{onsager1949statistical}. This is markedly different to the behavior of vortices in 3D fluids ~\cite{maurer1998local,henn2009emergence}. Onsager’s theory has provided some understanding of diverse classical quasi-2D systems such as turbulent soap films~\cite{Rutgers:1998uz}, guiding-center plasmas~\cite{Smith:1990cx}, self-gravitating systems~\cite{binney2011galactic}, and Jupiter’s Great Red Spot~\cite{young2017forward}. However, quantitative demonstration of point-vortex statistical mechanics is challenging; although the dynamics in two-dimensional classical fluids can lead to vortex cluster growth, these vortices are continuous and cannot be realistically modeled by discrete points~\cite{tabeling2002two,sommeria2001two}. Aware of this limitation, Onsager noted the model would be more realistic for 2D superfluids, where vortices are discrete, with circulations constrained to $\Gamma = \pm h/m$, where $h$ is Planck’s constant and $m$ is the mass of a superfluid particle. The physical realization of high-energy point-vortex clusters in any fluid system has however remained elusive. 

 \begin{figure}[!ht]
 \centering\includegraphics[width=\columnwidth, trim = 0cm 0cm 0cm 0cm, clip = false]{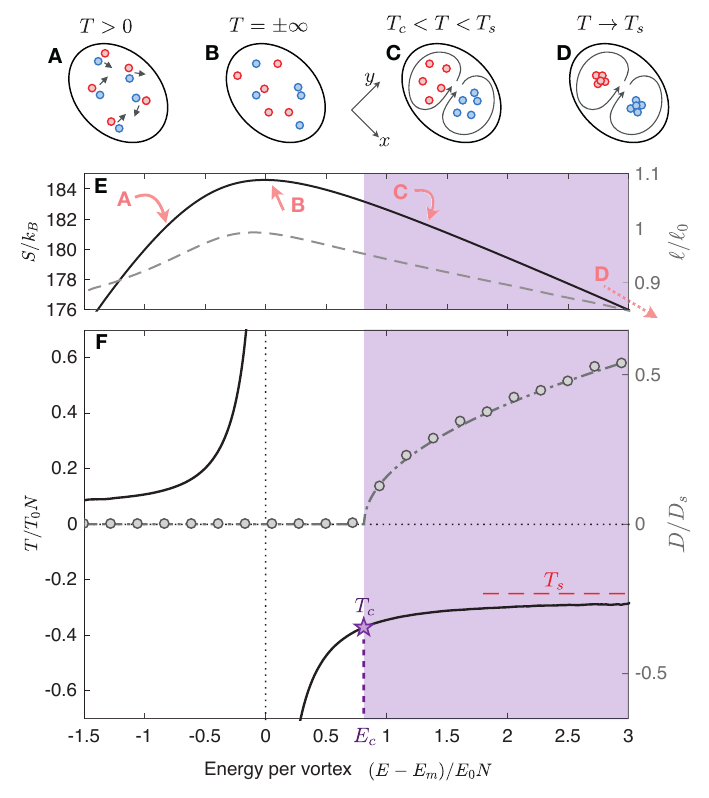}
	\caption{\textbf{Phases of point-vortex matter in a bounded domain.}
	(\textbf{A}) Small positive temperatures exhibit bound vortex-antivortex pairs. (\textbf{B}) As the vortex temperature $T\rightarrow \infty$ vortex positions become uncorrelated. (\textbf{C}) At high enough energies a clustering temperature $T_\textrm{c}$ is reached where giant Onsager vortex clusters form. (\textbf{D}) As $E\rightarrow \infty$ the clusters shrink to two separated points forming a supercondensate. (\textbf{E},\textbf{F}) Monte Carlo data for a neutral vortex gas, for the elliptical domain and vortex numbers ($N_+=N_-=N/2=9$) studied in the experiment. (\textbf{E}): Entropy $S$ and nearest-neighbour distance $\ell/\ell_0$, where $\ell_0=\sqrt{(ab/N)}$ for ellipse semi-major axis a and semi-minor axis b. (\textbf{F}) Temperature $T$ and the dipole moment (order parameter of the clustering transition) \small $D=N^{-1}|\sum\limits_{j}\textrm{sgn}(\Gamma_j)x_j|$\normalsize~as a fraction of the supercondensate limit $D_\textrm{s} \approx 0.47a$~\cite{refSupp}. Above the transition, $D\propto(E-E_c )^{1/2}$~\cite{refSupp} and a line of best fit yields the transition point $(E_c-E_\textrm{m})\approx 0.81E_0 N$, $T_\textrm{c}\approx-0.37T_0 N$ (purple star); the shaded region exhibits macroscopic vortex clusters. The red dashed line indicates the supercondensation limit, $E\rightarrow \infty$, $T\rightarrow T_s=-0.25T_0 N$. Here $E_0=\rho_0 \Gamma^2/4\pi$ and $T_0=E_0/k_\textrm{B}$, where $k_\textrm{B}$ is Boltzmann’s constant.}
	\label{DOSFig}
 \end{figure}

The incompressible kinetic energy  of an isolated 2D fluid containing N point vortices can be expressed in terms of the relative vortex positions~\cite{lin1941motion}. In an unbounded uniform fluid, it has the form

\begin{equation}
H = -\frac{\rho_0}{4\pi}\sum_{i \neq j}\Gamma_i\Gamma_j\textrm{ln}\left|\frac{\mathbf{r}_i - \mathbf{r}_j}{\xi}\right| \ ,
\label{eqn:PVH}
\end{equation}

\noindent where $\rho_0$ is the 2D fluid density, $\xi$ is a short-range cutoff scale, and $\Gamma_i$ is the circulation of a vortex at position $\mathbf{r}_i$;  the sign of $\Gamma_i$ indicates the direction of rotation. Onsager's key insight was that, because  Eq. 1 is determined by the positions $\mathbf{r}_i$, for a confined  fluid the available phase space becomes bounded by the area of the container~\cite{onsager1949statistical}. This property dramatically alters the system’s thermodynamic behavior.

\noindent The equilibrium phases of a neutral N-vortex system in a bounded elliptical region are shown schematically in Fig. 1, A-D. Thermodynamic equilibria maximize the entropy (Fig. 1E), given by $S(E)=k_\textrm{B} \textrm{ln}W(E)$ where the density of states \small $W(E) = \xi^{-2N}\int \prod\limits_{i}^{N} d^2\mathbf{r}_i \delta\left[E - H(\{\mathbf{r}_i\})\right]$ \normalsize measures the number of possible vortex configurations at a given energy E~\cite{refSupp};  $k_\textrm{B}$ is Boltzmann’s constant. The vortex temperature (Fig. 1F) is given by $T=(\partial S/\partial E)^{-1}$. The low energy, positive temperature phase ($T>0$) consists of bound vortex-antivortex pairs (Fig. 1A). As the energy increases these pairs unbind~\footnote{Adding a finite vortex-core size allows the Kosterlitz-Thouless transition to occur in this system~\cite{Kosterlitz1973}.}, increasing the average nearest-neighbour distance l (Fig. 1E), until the vortex distribution becomes completely disordered (Fig. 1B), marking the point of maximum entropy ($T=\infty$). However, owing to the bounded phase space, this point occurs at finite energy $E_\textrm{m}$; at still higher energies vortices reorder into same-sign clusters~\cite{onsager1949statistical,Montgomery:1974dra}, thus decreasing the entropy, and yielding negative absolute temperatures ($T<0$). At a sufficiently high energy the system undergoes a clustering transition ($T=T_\textrm{c}$)~\cite{yu2016theory}; here the vortices begin to polarize into two giant clusters of same-circulation vortices (Fig. 1C), whose structures are determined by the shape of the container. The major-axis projection of the dipole moment , \small $D=N^{-1} |\sum\limits_{j}\textrm{sgn}(\Gamma_j)x_j|$\normalsize, serves as an order parameter for the clustering transition [14]; below the transition $D=0$, whereas above the transition it begins to grow as $D\propto(E-E_c )^{1/2}$ (Fig. 1F)~\cite{yu2016theory}. Finally, in the so-called supercondensation limit $\xi \rightarrow 0$, $E\rightarrow \infty$, the clusters shrink to two separated points (Fig. 1D). Here the temperature approaches the limiting supercondensate temperature $T_s$, which is independent of geometry~\cite{Kraichnan:1975ku}, and the dipole moment approaches a maximum $D_\textrm{s}$, determined by the geometry. In a superfluid, the cutoff scale $\xi$ is provided by the superfluid healing length; vortex core repulsion at lengths $\sim \xi$ prevents the eventual point collapse at infinite energy by enforcing an upper energy limit with a minimum entropy~\cite{refSupp}.
 
 To physically realize this idealized model, the vortices must form a well-isolated subsystem and effectively decouple from the other fluid degrees of freedom. A large and uniform 2D Bose-Einstein condensate (BEC), near zero temperature with weak vortex-sound coupling, has been proposed as a suitable candidate system ~\cite{fetter1966vortices,billam2014onsager,simula2014emergence,salman2016long}.  Furthermore,  superfluids allow for vortex-antivortex annihilation, which favors the formation of Onsager vortices through evaporative heating~\cite{campbell1991statistics,simula2014emergence}, whereby annihilations remove low energy dipoles, thus increasing the remaining energy per vortex. However, although small transient clusters have been observed in BEC~\cite{Neely:2010gl,Neely:2013ef,Kwon:2016ek}, attempts to create Onsager’s vortex clusters have thus far been hindered by thermal dissipation and vortex losses at boundaries~\cite{Seo:2017ix}, which are enhanced by fluid inhomogeneities~\cite{groszek2016onsager}. This has prevented the experimental study of the full phase diagram of 2D vortex matter shown in Fig. 1.

Here we overcome these issues by working with a uniform planar  $^{87}$Rb BEC confined to an elliptical geometry~\cite{refSupp}. Although the BEC itself is three-dimensional, the vortex dynamics are two-dimensional owing to the large energy cost of vortex bending~\cite{rooney2011suppression,refSupp}. By engineering different stirring potentials, we can efficiently inject vortex configurations with minimal sound excitation~\cite{refSupp}. A high energy vortex configuration can be injected using a double-paddle stir, whereby two narrow potential barriers~\cite{white2012creation,Stagg:2014hl} are swept along the edges of the trap (Fig. 2A). Because of the broken symmetry of the ellipse, the maximum entropy state is a vortex dipole separated along the major axis~\cite{esler2015universal}. The stirring protocol is well mode-matched to this vorticity distribution, and we find the vortices rapidly organize into two Onsager vortex clusters (Fig. 2B).

\begin{figure}[t!]
\centering\includegraphics[width=\columnwidth, trim = 0cm 0cm 0cm 0cm, clip = false]{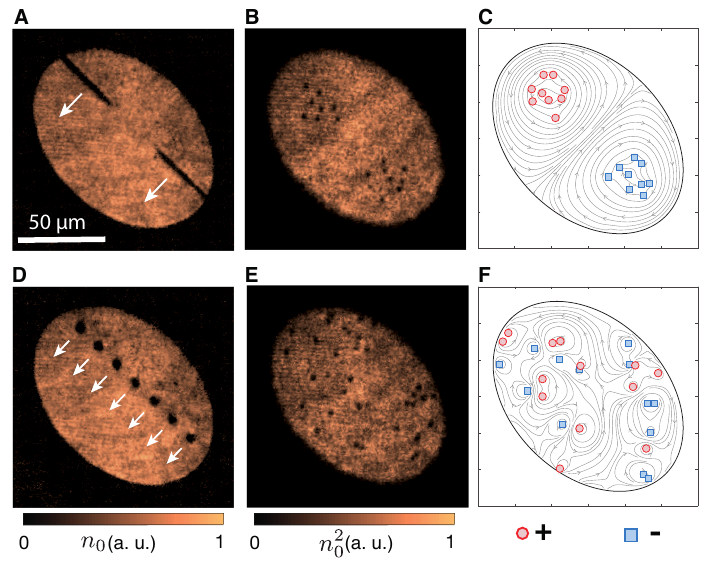}
	\caption{\textbf{Experimental vortex injection.} Experimental vortex injection. (\textbf{A}) Two large paddle potentials stir the BEC inducing large scale flow (In situ image, part-way through the stir). The white arrows indicate the direction of the stir (\textbf{B}) A 3 ms time-of-flight Faraday image directly after the paddle stir clearly resolves injected vortices~\cite{refSupp} localized into two clusters. (\textbf{C}) Simulation of the paddle stir showing velocity contours, with the location and circulations of the vortices demonstrating the injection of a clustered vortex dipole. (\textbf{D}-\textbf{F}) As for (\textbf{A}-\textbf{C}) but with a low-energy vortex distribution injected by a grid of narrow circular barriers.}
		\label{InjectionFig}
\end{figure}

\begin{figure}[t!]
	\centering{\includegraphics[width=\columnwidth]{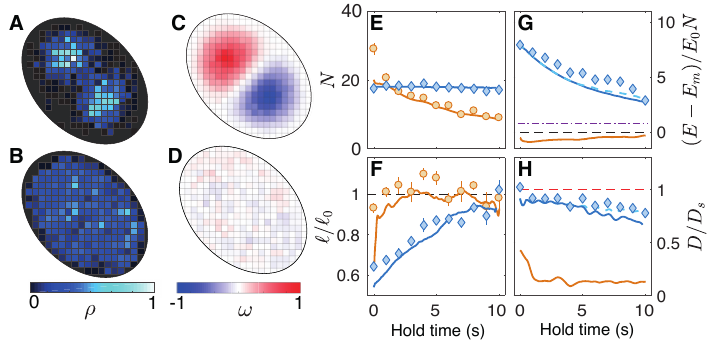}}
	\caption{\textbf{Evidence of Onsager vortex cluster metastability}. Experimental (unsigned) vortex density histograms $\rho = \sigma_+ + \sigma_-$ for (\textbf{A}) paddle and (\textbf{B}) grid stirs, respectively. The data are collected following hold times of $t=\{0,1,2...10\}$ seconds, with 10 samples at each time point (110 samples total). (\textbf{C}, \textbf{D}), Corresponding GPE simulation (signed) vorticity histograms $\omega=\sigma_+ - \sigma_-$ time averaged over 0 - 10 s. (\textbf{E}, \textbf{F}) Experimental average vortex number $\langle N \rangle$ and nearest-neighbor distance $\ell/\ell_0$ vs. hold time, where $\ell_0=\sqrt{0.89ab/N}$ is the expected value for a uniform distribution within the 89\% detection region of the $a$:$b$ ratio ellipse: paddle stir (blue diamonds) and grid stir (orange circles). GPE simulation results are shown as solid lines of the same color. (\textbf{G}) Point-vortex energy vs. time. Blue diamonds: experimental estimate. Blue solid line: exact point vortex energy from GPE. Blue dashed line: estimate from applying the experimental vortex classification method to the GPE data. The black horizontal dotted line indicates the energy of the state with $T=\pm \infty$ and the purple dash-dotted line indicates $T=T_\textrm{c}$. The orange line indicates the energy of the grid-stir simulation. (\textbf{H}) Dipole moment vs. time. Lines and markers are as in \textbf{G}. Red dashed line shows the supercondensate limit $D=D_\textrm{s}$. Simulations are averaged over 10 runs and a 1 s rolling time average.}
	\label{HistCompare}
\end{figure}

We contrast these results with the injection of a low energy configuration from sweeping a grid of smaller circular barriers through the BEC. Experimentally we find this results in a similar number of vortices (Figs. 2, D and E), but in a disordered distribution that can undergo evaporative heating~\cite{simula2014emergence,groszek2016onsager,johnstone2019evolution} (cf. Fig. 1B). Gross-Pitaevskii equation (GPE) simulations   quantitatively model both stirring methods and are compared in Figs. 2, C and F and Movies S1 and S2.

Although the detection of the vortex sign~\cite{Seo:2017ix,johnstone2019evolution} is possible~\cite{refSupp}, the clustered states are non-uniform equilibria, and their presence can also be confirmed from the (unsigned) vortex density $\rho = \sigma_+ + \sigma_-$, where $\sigma_+ (\sigma_-)$ denotes the distribution of positive (negative) vortices~\cite{refSupp}. Figure 3A displays a time-averaged position histogram, generated by measuring the experimental vortex positions at one-second intervals over ten seconds of hold time following injection. As expected for our elliptical geometry~\cite{refSupp}, the density shows two distinct persistent clusters separated along the major axis. The clusters remain distinguishable up to 9 s of hold time in individual frames (see Movie S3). By contrast, the grid stir in Fig. 3B shows a near uniform distribution of vortices consistent with an unclustered phase (Figs. 1, A and B). Figures 3, C and D show the corresponding (signed) density $\omega = \sigma_+ - \sigma_-$ from GPE simulations, showing polarized clusters for the paddle stir, contrasted with $\omega \approx 0$ for the grid stir. Figure 3E shows the total vortex number as a function of time for the two stirs in comparison with respective simulations. The vortex number for the paddle stir shows almost complete suppression of vortex decay over 10 s, indicating a strong spatial segregation of oppositely-signed vortices. In contrast, the grid stir loses 60\% of the vortices over this period to vortex annihilation and edge losses. Figure 3F plots the vortex nearest-neighbor distance $\ell/\ell_0$ (where $\ell/\ell_0\simeq1$ indicates a uniform distribution, cf. Fig. 1E). Although this quantity increases with time for the clustered state, indicating spreading of the clusters, it remains $<1$ for the entire 10 s duration. By contrast, for the grid stir $\ell/\ell_0$ stays quasi-constant and near unity, characteristic of a disordered state.

In the clustered phase our simulations demonstrate that vortex signs can be dependably inferred for $t \leq 5$~s from the experimental positions of the vortices relative to the minor axis of the ellipse [14]. From these data we can estimate the energy of the experimental vortex configurations as a function of time using the point-vortex model (including boundary effects [14]) and compare with GPE simulations, as shown in Fig. 3G. Despite a gradual decay of the energy, the system remains well within the negative temperature clustered region  for the entire 10 s hold time, equivalent to approximately 50 times the initial cluster turnover time of $\sim0.2$~s (see Movie S1). The decay is caused by a combination of the finite lifetime of the condensate ($\tau=28\pm2$~s), residual thermal fraction of $\sim30$\%, and residual non-uniform BEC density of $\sim6$\% RMS. This conclusion is supported by GPE simulations with phenomenological damping, which are in agreement with experimental observations (see Fig. 3G). The grid stir simulation shows a small increase in energy per vortex over the hold time, indicating that evaporative heating marginally prevails over thermal dissipation; annihilations manage to drive the system towards the negative temperature region, but not into the clustered phase.

Similarly, we may estimate the dipole moment, D and the vortex temperature T, which we compare with theoretical predictions. Figure 3H shows that the paddle stir exhibits a large dipole moment, with an average of $D/D_\textrm{s}\sim$81\% over the 10 second hold time. The experimental estimate agrees well with simulations for $t\leq5$ seconds, when opposite signed vortices remain completely segregated on opposite sides of the minor axis. By contrast, for the grid stir $D/D_\textrm{s}\sim1/\sqrt{N}$, consistent with an unclustered phase at finite $N$~\cite{yu2016theory,refSupp}. Finally, our Monte Carlo simulations~\cite{refSupp} show that the clustering transition occurs at a temperature $T_\textrm{c}\simeq -0.37T_0 N$, whereas supercondensation~\cite{yu2016theory} occurs at $T_s=-0.25T_0 N$ (see Fig. 1F). We estimate the final temperature from the point-vortex energy, finding $T_\textrm{exp}≃-0.28T_0 N$, consistent with the vortex system being in the clustered region of the phase diagram. 

Thermal friction is expected to play a major role in the damping of the Onsager vortex clusters~\cite{moon2015}. We experimentally investigated the role of an increased thermal component by injecting clusters for a range of smaller condensate fractions (i.e., higher BEC temperatures), while maintaining similar injected vortex number (Fig. S9). As shown in Fig. 4A, with decreasing condensate fraction we observe a reduction of the exponential decay time nearest-neighbor distance decay time  to the uniform value $\ell/\ell_0 \simeq 1$, obtained by empirical fits to the nearest-neighbour distance~\cite{refSupp}, (examples in Fig. 4B). These results indicate that with decreasing condensate fraction the vortices more rapidly approach a low energy, uncorrelated distribution;  cumulative vortex histograms for the largest and smallest condensate fractions (insets) also show diminished clustering with decreasing condensate fraction. Furthermore, the initial nearest-neighbor distance increases with decreasing condensate fraction (Fig. 4C), suggesting the injection of high energy clusters is less efficient with increased damping. These results suggest thermal dissipation is more important than losses to sound in our experiment; indeed, Gross-Pitaevski simulations without thermal damping (thus containing only losses from vortex-sound coupling) were found to support this conclusion as the clusters retained over 90\% of their initial energy~\cite{refSupp}.   Thermal friction may limit future experiments from observing the dynamic emergence of Onsager vortex clusters.

\begin{figure}[t!]
		\centering{\includegraphics[width=\columnwidth]{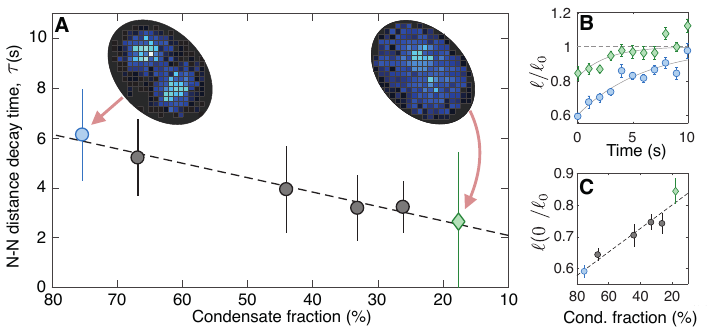}}
		\caption{\textbf{Cluster decay rate vs BEC fraction.}  (\textbf{A}) Decreasing condensate fraction results in more rapid cluster dissociation, indicated by decreasing nearest-neighbor (N-N) distance decay times as determined by exponential fits, (\textbf{B})  N-N distance decay for the largest (blue circles) and smallest (green diamonds) condensate fractions, with fits shown by full lines. Insets in (\textbf{A}) show time-averaged vortex density histograms accumulated over a 10 second hold for these cases, as in Fig. 3; see Fig. S7 for the full set of time-averaged histograms, and Fig. S8 for the histograms immediately following the stir. (\textbf{C}) The initial nearest-neighbor distance $\ell(0)$ increases with decreasing condensate fraction, indicating limitations in injecting high-vortex energy in the presence of thermal damping. The dashed line indicates a linear fit to the data.}
		\label{EnergyCompare}
\end{figure}

We note that, once achieved, the clustered phase is remarkably robust to dissipation, contrary to the conventional wisdom for negative temperature states. Meanwhile, the evaporative heating mechanism appears to be more fragile, inhibited by modest dissipation. Nonetheless, a systematic study of the clustering transition and its emergence from quantum turbulence~\cite{billam2014onsager,simula2014emergence,salman2016long} appears within reach, if further reduction of thermal dissipation can be achieved. The precise control of the trapping potential in our experiment enables a broad range of stirring and trapping configurations, opening the door to further studies of the vortex clustering phase transition~\cite{billam2014onsager,simula2014emergence,yu2016theory}, and of fully developed quantum turbulence confined to two dimensions. Emerging tools for precision characterization, including vortex circulation detection~\cite{Seo:2017ix}, momentum spectroscopy~\cite{reeves2014signatures}, and correlation functions~\cite{white2012creation,Skaugen:2016hq}, can be expected to provide further insights into the role of coherent structures in 2D vortex matter.

We note that negative absolute temperature vortex states in a different regime, along with signatures of evaporative heating by vortex-antivortex annihilation, were independently observed in~\cite{Johnstone2018}.\\


\noindent\textbf{Acknowledgements:} 
We thank B. P. Anderson, L. A. Williamson, P. B. Blakie, K. Helmerson, T. Simula, S. P. Johnstone, A. J. Groszek, M. Cawte and G. J. Milburn for useful discussions.\\
\textbf{Funding}: This research was partially supported by the Australian Research Council through the ARC Centre of Excellence for Engineered Quantum Systems (project numbers CE1101013, CE170100009), the ARC Centre of Excellence in Future Low-Energy Electronics Technologies (project number CE170100039), ARC Discovery Project DP160102085, and funded by the Australian Government. Further support was provided by the Dodd-Walls Centre for Photonic and Quantum Technologies, the Royal Society of New Zealand Marsden Fund (contract UOO1726). T.A.B. and G.G. acknowledge the support of an Australian Government Research and Training Program Scholarship.

%

\clearpage \newpage
\section*{Materials and Methods}

\noindent\textbf{Optically configured Bose-Einstein condensates (BECs).} Our experimental apparatus consists of a $^{87}$Rb BEC confined in a red-detuned laser sheet, providing harmonic trapping in the vertical $z$ dimension with frequency $\omega_z = 2\pi \times 108$ Hz.  The trapping in the $x$-$y$ plane can be arbitrarily configured via direct projection of blue-detuned light which is  patterned with a digital micromirror device (DMD)~\cite{gauthier2016direct}.  The BEC is formed using a hybrid optical and magnetic trapping technique~\cite{lin2009rapid}.  We initially evaporate in a hybrid trap produced from a single, radially symmetric 95~$\mu$m waist 1064~nm red-detuned Gaussian beam and weak quadrupole magnetic field~\cite{gauthier2016direct}. Before reaching the BEC critical temperature, we transfer the atoms to a 1064~nm red-detuned Gaussian sheet and simultaneously ramp the magnetic field to approximately zero. Optical evaporation over four seconds produces BECs of up to  $N_{\textrm{c}} = 3 \times 10^6$ atoms in the approximately azimuthally symmetric harmonic optical trap with $\{\omega_x,\omega_y,\omega_z\} = 2 \pi \times \{6.8,6.4,360\}$~Hz. In the final second of evaporation, the intensity of the 532~nm light illuminating the DMD is linearly ramped, resulting in a peak potential value of $10\mu$, where $\mu = k_B \cdot 22$~nK is the chemical potential, producing a highly-oblate configured BEC with $N_{\textrm{c}} \sim 2.2 \times 10^6$ and 67(3)\% condensate fraction in a hard-walled elliptical trap with major and minor axes $\{2a,2b\} = \{125,85\}~\mu$m. With the optical trapping beams held on, we levitate the cloud against gravity by ramping on an unbalanced quadrupole magnetic field, which additionally results in a 80~G DC residual magnetic field in the vertical direction. Simultaneously, we reduce the sheet trapping power resulting in the final trap frequencies $\{\omega_x,\omega_y,\omega_z\} \sim 2 \pi \times \{1.8,1.6,108\}$~Hz, and trap depth of $\sim90$~nK.  We also reduce the DMD pattern depth to $\sim5\mu$. Combined with the hard-walled confinement of the DMD, this results in an approximately uniform atom distribution with a calculated vertical Thomas-Fermi diameter of $6~\mu$m and a healing length of $\xi \sim 500$~nm at the center of the trap (average $\xi \sim 530$~nm).

We measure the BEC lifetime in this trap to be $\sim 28$~s, which is shorter than the vacuum-limited lifetime of $\sim 60$~s.  This suggests that scattering from the optical trap is a source of atom loss.   We expect that a trap based on blue-detuned light would reduce this loss, and along with  increased condensate fraction could potentially increase the lifetime of the vortex clusters.\\

\noindent\textbf{Obstacle stirring protocols.} The paddle and grid obstacles are formed using the DMD.  To dynamically alter the potential we upload multiple frames to the DMD, with the initial frame being the empty elliptical trap. Elliptically-shaped paddles, with a major and minor axis of $85~\mu$m and $2~\mu$m respectively, are then swept through the BEC at constant velocity. The paddle stirs are defined by a set of 250 frames and the barriers start slightly outside the trap edge, with the paddles intersecting the edges of the elliptical trap at their midpoints.  A $150~\mu \textrm{m}~\textrm{s}^{-1}$ sweep ($\sim0.1 c $, where the speed of sound $c \sim 1290~\mu \textrm{m}~\textrm{s}^{-1}$) is utilized for the $\{2a,2b\} = \{120,85\}~\mu\textrm{m}$ trap, which results in a sweep time of $580$~ms.  Sequential paddle positions are separated by $\sim350$~nm, resulting in sufficiently smooth translation. After crossing the halfway point, the paddles are linearly ramped to zero intensity by reducing the major and minor axes widths to zero DMD pixels. For the grid case, an array of seven $4.5~\mu\textrm{m}$ diameter barriers were swept at the increased velocity of $390~\mu\textrm{m}~\textrm{s}^{-1}$, due to a higher critical velocity for vortex shedding. The barriers heights were then linearly ramped to zero after crossing the halfway point.\\

\noindent\textbf{BEC imaging and vortex detection.} For darkground Faraday imaging~\cite{ccbradley} we utilize light detuned by 220~MHz from the $^{87}$Rb $|F=1\rangle\rightarrow|F'=2\rangle$ transition in a 80~G magnetic field with $52.6\times$ magnification. This results in images with a measured resolution of 960(80)~nm  FWHM at 780~nm illumination~\cite{gauthier2016direct}. For the small phase shifts imparted by our vertically-thin cloud, raw Faraday images return a signal $\propto n_0^2$ which improves vortex visibility through exaggerating density fluctuations. The density can be determined through post-processing. The $\xi\sim500$~nm healing length results in poor vortex visibility \emph{in situ}, see Figs.~2, {A and D}.  However, Faraday imaging combined with a short 3~ms time of flight (TOF), where the optical beams are suddenly turned off, but the levitation field is held on, improves the vortex visibility significantly, while the column density is otherwise essentially unchanged, see Figs.~2, {B and E}. After masking the image with the elliptical pattern, vortices are detected automatically using a Gaussian blob vortex image processing algorithm~\cite{rakonjac2016measuring} that examines connected regions of a thresholded background-subtracted image, see Fig.~\ref{SuppDetection}{}. We restrict detection of vortices to the inner 89\% of the ellipse area to avoid spurious detections near the condensate edge.\\

\noindent\textbf{Effective 2D theory.} The oblate atomic BEC can be modeled by the Gross-Pitaevskii equation (GPE)
\setcounter{equation}{0} 
\begin{equation}
i\hbar \partial_t \Phi(\mathbf{r},t ) = \left[- \frac{\hbar^2 \nabla^2}{2m} + V(\mathbf{r},t) + g |\Phi|^2 -\mu \right] \Phi(\mathbf{r}, t),
\label{eqn:3DGPE}
\end{equation}

\noindent with interaction parameter $g = 4\pi\hbar^2 a_s /m$ for s-wave scattering length $a_s$, atomic mass $m$, and
chemical potential $\mu$. The trapping potential $V(\mathbf{r})$ can be modeled as the sum of harmonic vertical confinement and a hard-wall DMD potential in the $x$-$y$ plane
\begin{equation}
V(\mathbf{r}) =  V(z) + V(x,y) = \frac{1}{2}m \omega_z^2 z^2 + V_0\; \Theta \left(\frac{x^2}{a^2} + \frac{y^2}{b^2}  - 1\right),
\end{equation}

\noindent where $\Theta(x)$ is the Heaviside function and $V_0 \gg \mu$. In the Thomas-Fermi approximation, the chemical potential of the ground state is then
\begin{equation}
\mu = \frac{1}{2}\left(\frac{3 g N_{\textrm{c}} (m \omega_z^2)^{1/2}}{2\pi a b} \right)^{2/3}, \,\text{i.e.}\,, \quad \mu \sim N_{\textrm{c}}^{2/3 }.\label{eqn:mu3D}
\end{equation}

To obtain an effective 2D theory, in the usual way we write $\Phi(\mathbf{r},t) = \phi(x,y,t) \chi(z)$. However, as the confinement in $z$ is not strong enough to confine the wavefunction to the harmonic oscillator ground state, $\chi(z)$ is instead approximated by the Thomas-Fermi profile: $\chi(z) = \left( 3 g/4 \mu l_z\right)^{1/2} \sqrt{(\mu - V(z)/g},$ where $l_z = (2\mu/m\omega_z^2)^{1/2}$ is the axial Thomas Fermi radius, such that $\chi(z)$ satisfies $\int dz\;  |\chi(z)|^2 = 1$ and $\int d^2 \mathbf{x} \; |\phi(x,y)|^2 = N_{\textrm{c}}$. Multiplying by $\chi^*(z)$ and integrating over $z$ yields an effective 2D equation of motion

\begin{equation}
i\hbar\partial_t \phi = \left[- \frac{\hbar^2}{2m}(\partial_x^2 + \partial_y^2) + V(x,y)  + g_2 |\phi|^2 - \mu_{2D} \right] \phi,
\label{eqn:2DGPE}
\end{equation}

\noindent where the effective 2D interaction strength is $g_2 = {3g}/{5l_z}$ and $\mu_{2D} = 4 \mu/5.$ For a BEC of $N_{\textrm{c}} = 2.25 \times 10^{6}$ atoms in the $\{2a,2b\} = \{120,85\}~\mu$m trap, we obtain $\mu_{2D}/k_B  = n_0 g_2/k_B = 19.63 \text{ nK}$, $\xi = \hbar/\sqrt{ m \mu_{2D}}  \approx 0.533 \; \mu \textrm{m}$ and $c  = \sqrt{\mu_{2D}/m} \approx 1370$ $\mu \textrm{m~s}^{-1}.$ Values for the systems with lower condensate fraction or larger trap size are obtained by a simple scaling. Scaling the condensate number $N_{\textrm{c}}  \rightarrow \alpha N_{\textrm{c}}$  gives $\xi \rightarrow \alpha^{-1/3}\xi$ and $c \rightarrow \alpha^{1/3} c $ while scaling the trap $ \{a,b\}  \rightarrow \lambda \{a,b \}$ yields $\xi \rightarrow \lambda^{2/3} \xi$ and $c \rightarrow \lambda^{-2/3} c$.\\

\noindent\textbf{Dynamical modeling.} We model the dynamical evolution of the experiment using a phenomenologically damped GPE to account for energy and atom losses~\cite{choi1998phenomenological}. Eq.~(\ref{eqn:2DGPE}) becomes
\begin{equation}
i \hbar\partial_t \phi = (1-i \gamma)\left [-\frac{\hbar^2\nabla^2}{2m} + V(x,y,t) + g_2|\phi|^2 - \mu(t) \right]\phi,
\label{eqn:dGPE}
\end{equation}

\noindent where $\gamma$ is the dissipation coefficient. Up to a  noise term, Eq.~(\ref{eqn:dGPE}) is equivalent to the simple growth stochastic Gross Pitaevksii equation (SGPE), a microscopically derived model of atomic BECs that incorporates dissipation due to interactions with a thermal component~\cite{rooney2012stochastic}. The exponential decay of the atom number, $N_{\textrm{c}}(t) = N_{\textrm{c}}(0) \exp(-t/\tau)$, for the decay constant $\tau \approx 28 \pm 2$ s, is incorporated via a time-dependent chemical potential $\mu(t)$. From Eq.~(\ref{eqn:mu3D}) this gives  $\mu(t) =\mu_{2D} \exp(-2 t/3\tau)$. Empirically we find that the experimental data for the paddle stir are well matched by numerical simulations with a dissipation coefficient of $\gamma = 6.0 \times 10^{-4}$. We find a slightly larger phenomenological dissipation coefficient is required for the grid stir, $\gamma = 8.5 \times 10^{-4}$. We attribute this to the increased sound production for this case (see below). The total external potential is modeled as a combination of a stationary trap and time-dependent stirring obstacles: $V(x,y,t) = V_{\textrm{trap}}(x,y) + V_{\textrm{ob}}(x,y,t)$. The stationary component of the trap includes the optical dipole trap in the $x$-$y$ plane, and binary DMD pattern convolved with the previously measured point spread function of the optical system~\cite{gauthier2016direct}. The stirring obstacles are modeled by steep-walled hyperbolic tangent functions, which increase to the maximum on the scale of the healing length. The numerical simulations were performed using XMDS2~\cite{dennis2013xmds}.

Note that while the damped GPE simulations are a reasonable approximation at high condensate fractions, as shown in Fig.~3, full stochastic simulations would likely be required for the high-temperature and low condensate-fraction conditions of Figs.~4 and \ref{SuppHist}. As the phenomenological damping parameters are  determined \emph{a posteriori}, we did not simulate these cases.

We also performed undamped GPE simulations to obtain an upper estimate of the sound produced from the stirring procedures. Using the standard Helmholtz decomposition on the kinetic energy~\cite{nore1997kolmogorov}, we find the amount of sound produced is quite small: for the paddle stir it is $\sim 1.5\%$ of the total kinetic energy at the end of the stir and $< 5\%$ at the end of the hold time (the increase is from vortices radiating sound as they accelerate). For the grid stir it is $\sim 8 \%$ at the end of the stir and $<13\%$ at the end of the hold time. As the undamped simulations include sound radiation effects but not thermal damping, we also compared damped and undamped simulations to distinguish between the effects of thermal and acoustic losses. In undamped simulations, the  (incompressible) vortex subsystem retained 90\% of its energy over the 10~s period, compared to only $\sim$40\% in the damped case, indicating that thermal losses are more significant than losses from radiation to sound.

While vortex bending becomes highly suppressed in 3D oblate harmonic traps (\emph{28}), we also performed 3D GPE simulations of Eq.~(\ref{eqn:dGPE}), shown in Fig.~\ref{Supp3DGPE}. Little to no vortex bending is observed, further justifying a 2D treatment of the system.\\

\noindent\textbf{Point-vortex energy.} The point-vortex Hamiltonian can be constructed for any simply connected domain using a conformal map to the unit disc combined with the method of images. Under a conformal map $\zeta = f(z)$, which derives the vortex motion in the domain $z \in \Omega$ from that in the domain $\zeta \in \mathcal{D}$, the Hamiltonians are related via~\cite{newton2013nvortexproblem}

\begin{equation}
H_{\Omega}(z_1, \dots, z_N)  =  H_{\mathcal{D}}(\zeta_1,\dots,\zeta_N) - \textstyle{\sum_{j=1}^N \kappa_j^2 \textrm{ln} \left|\frac{d\zeta}{dz}\right|_{z = z_j}}, \label{eqn:Htransform}
\end{equation}

\noindent where $z_j = (x_j + i y_j)/\xi$ is the complex position of the $j$th vortex and $\kappa_{i}=\pm 1$. Note that the energy in Eq.~(\ref{eqn:Htransform}) is dimensionless; physical energies can be obtained by multiplying by the energy unit $E_0=\rho_0 \Gamma^2 /4\pi$, where $\rho_0$ is the fluid density and $\Gamma = h/m$ is the unit of circulation for atomic mass $m$. 
If the map $\zeta = f(z)$ transforms a (simply connected) domain $\Omega$ to the unit disk  $\mathcal{D} = \left\{\zeta \in \mathbb{C} \,|\, |\zeta| \leq 1 \right \}$, Eq.~(\ref{eqn:Htransform}) gives~\cite{newton2013nvortexproblem}

\begin{equation}
H_{\Omega} =  \textstyle - \sum_j  \kappa_j^2 \textrm{ln} \left| \frac{\zeta'(z_j)}{1 - |\zeta_j|^2} \right| -  \sum_{j,k}' \kappa_j \kappa_k \textrm{ln} \left | \frac{\zeta_j - \zeta_k}{1 - \zeta_j \zeta_k^*}\right|  \label{eqn:ConformalEnergy},
\end{equation}

\noindent where $\zeta_j \equiv f(z_j)$ and the prime on the second sum indicates the exclusion of the term $j=k$. The domain of the ellipse interior, $ \Omega = \left\{ z \in \mathbb{C} \,|\, \Re(z)^2/a^2 + \Im(z)^2/b^2 \leq 1 \right\}$ is mapped to the unit disk by the conformal map~\cite{kober1957dictionary}

\begin{equation}
\zeta = f(z) = \sqrt{k} \; \mathrm{sn} \left( \frac{2K(k)}{\pi} \sin^{-1} \left( \frac{z}{\sqrt{a^2 - b^2}} \right) \,;\, k \right).
\label{eqn:ConformalMap}
\end{equation}

\noindent Here $\mathrm{sn}\,(z\, ; k)$ is the Jacobi elliptic sine function,
$K(k)$ is the complete elliptic integral of the first kind, and $k$ is the elliptical modulus, given by

\begin{equation}
k^2 = 16 \rho \prod_{n=1}^{\infty} \left(  \frac{1 + \rho^{2n}}{ 1 + \rho^{2n-1}}\right)^8,
\label{eqn:ConformalEnergy2}
\end{equation}

\noindent where $\rho = (a-b)^2/ (a+b)^2$. From Eqs.~(\ref{eqn:ConformalEnergy}), (\ref{eqn:ConformalMap}), and~(\ref{eqn:ConformalEnergy2}), the dynamics of point vortices in the ellipse can be calculated from Hamilton's equations as
\begin{equation}
    \kappa_j\dot{z_j} = -2i\,\partial H_\Omega/\partial z_j^*.
    \label{eqn:HamiltonPVE}
\end{equation}

\noindent\textbf{Monte Carlo sampling.} To generate the curves in Fig.~1{}, we generate $10^9$ uniformly random, neutral configurations of $N_{\pm}=9$ vortices within the ellipse, and calculate the dipole moment, mean nearest-neighbor distance and energy for each state. We then bin the energy samples to approximate the density of states, $W(E)= \xi^{-2N}\int \prod^{N}_{i=1} d^2 \mathbf{r}_i \delta(E-H(\{\mathbf{r}_i\}))$, which determines the entropy $S = k_B \log W$, where $k_B$ is Boltzmann's constant and $T = (\partial S/\partial E)^{-1}$. The nearest neighbor distances are binned according to their corresponding energies. To extract the dipole moment, following Ref.~\cite{ashbee2014dynamics, esler2013} we create a histogram of the energy and dipole moment, $p(E,D)$. The dipole moment $D(E)$ is determined by the peak of the distribution at each energy. 
Note that the mean dipole moment is also an indicator of the transition (as used in Fig.~3)~(\emph{16}), particularly at energies far above the transition where fluctuations are suppressed. However, below the transition the uncorrelated vortex positions yield $D\sim 1/\sqrt{N}$ (cf. Fig.~3{H}). For small $N$ this complicates the extraction of the critical scaling for $D(E)$ (see ``Onset of clustering" below).

Note that the energy in Eq.~(\ref{eqn:ConformalEnergy}) is defined up to an arbitrary additive constant.  Throughout this work we express energy relative to the maximum entropy point $E=E_\textrm{m}$ (where $T=\infty$) determined from the Monte Carlo simulations. \\
 
 \noindent\textbf{Onset of clustering.}
 \sloppy
For an incompressible flow, one can introduce a stream function  $\psi$,  connected to the vorticity $\omega(\mathbf{r})=\sigma_{+}(\mathbf{r})-\sigma_{-}(\mathbf{r})$ via
$-\nabla^2 \psi =4 \pi \omega$. The maximum entropy states for a system containing a large number of vortices in a bounded domain $\Omega$ can be described by the  self-consistent mean-field equation~(\emph{2})
\begin{equation}
\label{sc}
-\nabla^2 \psi= 4 \pi \left[\frac{n_0}{2}\exp(-\tilde\beta \psi)- \frac{n_0}{2}\exp(\tilde\beta \psi)\right],
\end{equation}
\noindent where $n_0=2/A$ is the normalized vortex number density inside the area $A$ , and $\tilde\beta\equiv (E_0 N/2k_BT)$ is the inverse temperature in the natural energy units of the vortex system.
Linearizing Eq.~(\ref{sc}) around the uniform state of vortices $\psi=0$, the fluctuation $\delta \psi$ satisfies
\begin{equation}
\label{LE}
(\nabla^2 + \lambda) \delta \psi=0,
\end{equation}

\noindent with the Dirichlet boundary condition $\delta \psi (\mathbf{r} \in \partial \Omega)=0$,  here $\lambda=-4 \pi \tilde\beta n_0$.
The onset of the vortex clustering (purple star in Fig.~1{F}) occurs if  Eq.~(\ref{LE}) has nonzero solutions to the eigenvalue problem of  the Laplacian  operator in the elliptical domain~(\emph{16}).
In terms of elliptical coordinates  Eq.~(\ref{LE})  becomes Mathieu's equation.
The most relevant eigenvalue associated with the transition is
$\lambda = 4 h^2/(a^2-b^2)$, where $h$ is the first positive root of  the modified Mathieu function   $\textrm{Mc1}(m, R, h)$ with $m=1$ and $ R= \textrm {tanh}^{-1}(b/a)$.
The transition happens at  $\tilde\beta=\tilde\beta_c = -\lambda/(4 \pi n_0)\simeq -1.614$, giving $T_\textrm{c}^{(mf)}=(k_{\textrm B}\tilde\beta_c)^{-1} E_0 N/2\simeq -0.31 T_0 N$, with $T_0=E_0/k_{\textrm B}$. Note that finite-$N$ effects are expected to lower the transition  energy and temperature~(\emph{16}).

In terms of vorticity, the macroscopic dipole moment can be expressed as 
\begin{equation}
D=|\mathbf{D}|=\frac{1}{2}\left|\int d^2\mathbf{r} \: \mathbf{r}  \omega  \right|.
\end{equation}
Due to the symmetry of the most relevant mode, $D_y=0$ and hence $D=|D_x|$. Near the transition $D \sim D_0 \left[(E-E_c)/(E_0 N)\right]^{\nu}$, where the critical exponent $\nu=1/2$ and the coefficient $D_0 \simeq 0.46 D_\textrm{s}$. Note that the exponent $\nu$ is universal while the coefficient $D_0$ depends on the geometry of the domain~(\emph{16}).

As the mean-field theory outlined above predicts $D \sim (E-E_c)^{1/2}$ near the clustering transition, we may use this scaling to extract the clustering energy $E_c$ and clustering temperature $T_\textrm{c}$ for the $N=18$ Monte Carlo simulations. We find the data are well described by the line of best fit
\begin{eqnarray}
D(E) &=&0  \quad \text{for} \quad E<E_c, \\
D(E) &=& D_0 [(E-E_c)/(E_0 N)]^{1/2} \quad \text{for} \quad E>E_c,  \end{eqnarray}
which yields $E_c - E_\textrm{m} \approx 0.81 E_0 N$ and $D_0/D_\textrm{s}\approx 0.37$, close to the mean field prediction of $D_0/D_\textrm{s}\approx 0.46$. The transition temperature can then be read off from the entropy function at $E=E_c$, giving $T_\textrm{c}(E_c) \approx -0.37 T_0 N$, consistent with the analytical mean field prediction $T_\textrm{c}^{(mf)} \simeq -0.31 T_0 N$, and the expected reduction of $T_\textrm{c}$ at finite $N$.\\

\noindent\textbf{Upper bound of the dipole moment.} Considering  two point vortices with opposite circulations in the $\{2a,2b\} = \{120,85\}~\mu\textrm{m}$ elliptical domain,  the mechanical equilibrium condition reads

\begin{equation}
\label{EL}
0=\kappa_{i} \frac{d x_i }{dt}=\frac{\partial H_{\Omega}}{ \partial y_i};  \quad
0=\kappa_{i} \frac{d y_i }{dt}=-\frac{\partial  H_{\Omega}}{ \partial x_i},
\end{equation}

\noindent where $i=1,2$, giving that $y_1=y_2=0$ and $x_1=-x_2=d$ is the unique stationary point where the forces on each vortex due to the other vortex and the image vortices all cancel. Numerically solving Eq.~(\ref{EL}), we find that $d\simeq 0.47a$. The value of $d$, which depends on the geometry of the domain, sets the upper bound of the average dipole moment in the supercondensate limit $\xi \rightarrow 0$, $E \rightarrow \infty$ for a neutral $N$-vortex system, where the vortices form two tight clusters near the fixed points ($D_\textrm{s} = d$). In the presence of a finite core $0 < \xi \ll \{a,b\}$, for our $N=18$ vortex system the highest energy state can be estimated by placing the vortices and antivortices each in a regular square array of $9$ vortices, centered on the equilibrium point $x = \pm 0.47a$, $y=0$. The dipole moment for such a state is the same as the supercondensation limit $D_\textrm{s}$. A crude but conservative estimate for the upper energy limit can be obtained as follows. Assume hard-core repulsion between vortices separated by $\sim10\xi$ (at a distance $5\xi$ from an isolated line vortex the superfluid density recovers to 98\% of the background value). For our $N=18$ vortex system, this yields an upper energy limit $(E-E_\textrm{m})/E_0 N \sim 12$, considerably larger than the largest energy achieved in the experiment ($(E-E_\textrm{m})/E_0 N \sim 8$), and greatly exceeding the largest energies presented for the Monte Carlo simulations $(E-E_\textrm{m})/E_0 N = 3$, indicating that the core repulsion is not significant in our experiment.\\

\noindent\textbf{Non-uniformity of clustered vortex states.} As noted in the main text, a key feature of the clustered states is that they are non-uniform, not only in the vorticity field $\omega(\mathbf{r})  = \sigma_+(\mathbf{r}) - \sigma_-(\mathbf{r})$, but also in the vortex density, $\rho(\mathbf{r})  = \sigma_+(\mathbf{r}) + \sigma_-(\mathbf{r})$. This contrasts with unclustered states, which exhibit uniform vortex density. 

Figure~3A of the manuscript shows that the vortex density histogram remains non-uniform for the entire hold time for the paddle stir. This can only happen if the vortex energy is sufficiently high, since at low energies vortices are free to roam throughout the entire system. The complete absence of vortex number decay for the paddle stir (Fig.~3E of manuscript) further supports this conclusion, as this requires the spatial segregation of positive and negative vortices which can only occur in a high energy configuration. By contrast, the grid stir exhibits significant vortex decay (because vortex-antivortex pairs are present), Fig.~3E, and the density is uniform, Fig.~3C.

To strengthen the argument above, we also consider average vortex densities at different energies under point vortex evolution via Eqs.~(\ref{eqn:ConformalEnergy}--\ref{eqn:HamiltonPVE}). In Fig.~\ref{fig:timeAverageDensity} we show time-averaged vortex densities produced from 2D point vortex dynamics, using a sample from the paddle experiment for the initial vortex positions. The energy can be altered by changing some of the vortex signs (while maintaining $N_+ = N_- = N/2$) for the same vortex position data. Assuming the opposite charges are completely segregated (far left) yields a histogram consistent with the experimental observations, whereas random charges (far right) instead yields constant density, as is observed for the low-energy grid stir. Only the left two panels resemble the experimental data in Fig.~3A. The simulations producing Fig.~\ref{fig:timeAverageDensity} did not contain any damping. However, dissipation would further smear the distributions, making the argument in favor of near complete clustering stronger. 

While the Bragg-scattering procedure for sign detection~(\emph{26,32}) is possible for our system, the non-uniform nature of the clustered states means we do not require sign detection for the majority of the analysis presented. Nonetheless, Bragg-scattering data was used to experimentally ensure the paddle experiment does indeed initially inject single-sign clusters. An example is shown in Fig~\ref{fig:Bragg}.

In order to determine the most likely vortex configuration, we calculate the standard deviation (standard error) between our experimentally observed Bragg-scattering differential signal, Fig.~\ref{fig:Bragg}C, and a simulated Bragg differential signal. The simulated signal starts with a point-vortex velocity field based on the experimental vortex positions, Fig.~\ref{fig:Bragg}B, where the circulations of the vortices can be iterated continuously between $\pm \Gamma$. By using this velocity field, along with the experimentally measured Bragg-scattering response function and total BEC density, a simulated differential signal can be generated for any configuration of vortex circulations. A steepest-decent method is used to minimize the standard error between the measured differential signal and simulated profile, with the initial guess being zero circulation for all vortices. The algorithm determines that a fully polarized vortex distribution with two same-sign clusters minimizes the standard error. To confirm that the steepest-decent converges to the global minimum, Fig.~\ref{fig:Bragg}D displays a histogram of all the possible $2^{16}$ vortex configurations vs. standard error. While this second approach confirms the configuration for which the standard error is minimized, it has the disadvantage of needing to iterate through all $2^N$ possible vortex-sign permutations.\\

\noindent\textbf{Nearest-neighbor distance and energy decay for varying BEC fraction and density.} For investigating vortex cluster energy damping as a function of BEC fraction, we reduce the depth of the evaporative cooling ramp, leading to an increased temperature and decreased condensate fraction.  During the levitation procedure, we find loss of thermal atoms for the hotter conditions which occurs at a rate inefficient for continued evaporation, due to the reduction in the optical dipole intensity and corresponding reduction in trap depth. This results in an approximately constant $N_{\textrm{tot}} \sim 3.3 \times10^6$ atoms in the final potential, while the final temperature and condensate fraction vary. The full range of temperatures and condensate fractions utilized were $T = \{$23(1), 24(2), 27(1), 30(2), 31(2), 32.7(2)$\}$~nK and $N_{\textrm{c}}/N_{\textrm{tot}} = \{$75.3(4), 67(3), 44(1), 33(1), 26(1), 18(2)$\}\%$, respectively.

Gross-Pitaevski equation (GPE) simulations have previously shown that increasing non-uniformity in the density of the condensate inhibits the dynamic formation of Onsager vortices~(\emph{27}).  To determine the sensitivity of our experiment to non-uniform density, we have increased the variation in the  density of our BEC before performing the paddle stir.  This is achieved by increasing the size of our trap while maintaining its aspect ratio, which increases the relative contribution of the residual harmonic optical trap from the optical dipole sheet potential to the 2D confinement. We thus examined three different trap sizes,  $\{2a,2b\} = \{125,85\}~\mu\textrm{m};\{140,100\}~\mu\textrm{m}; \{160,115\}~\mu\textrm{m}$ with RMS density variation of $\Delta n_0 =  \{6.2\%,$ $8.1\%,$ $10.9\% \}$, respectively. For the larger elliptical traps, the paddle sizes are proportionately scaled. A $150~\mu \textrm{m}~\textrm{s}^{-1}$ paddle stir is maintained for the $\{140,100\}~\mu\textrm{m}$ trap, but a $136~\mu\textrm{m}~\textrm{s}^{-1}$ velocity is used to produce a similar number of vortices for the $\{160,115\}~\mu\textrm{m}$ trap. For the $\{140,100\}~\mu\textrm{m}$ trap, the temperature and condensate fraction was $T =29(2)$~nK and $N_{\textrm{c}}/N_{\textrm{tot}} = 75.1(3)\%$, while the $\{160,115\}~\mu\textrm{m}$ trap had $T =36(4)$~nK and $N_{\textrm{c}}/N_{\textrm{tot}}  = 71(1)\%$.

As the total atom number $N_{\textrm{tot}}$ is approximately constant for all conditions, varying the BEC fraction and trap size leads to varying healing lengths, which we scale appropriately for calculating the vortex energy and nearest-neighbor distance. We furthermore determine the density of states $W(E)$ for $N_{\pm} = 9$ vortices to determine the peak value for each healing length, and shift the energies as described in previous sections. 

For the nearest-neighbor distance decay shown in Figs.~4 and \ref{SuppNNDecay}, we fit an empirical exponential decay function $\ell(t)/\ell_0 = ae^{-t/\tau}+1$, where the limiting value $\ell/\ell_0 \simeq 1$ is expected for uniformly distributed vortices. The resulting variation in nearest-neighbor distance decay times when varying the density is shown in Fig.~\ref{SuppNNDecay}A, which, in contrast to Fig.~4, shows little variation in the decay rate, and reduced variation in the initial nearest-neighbor distance, Fig.~\ref{SuppNNDecay}C.

We also apply the energy estimation procedure to the data; we fit the energy decay with the empirical function $E(t)/N = ae^{-t/\tau}+h_0$, where $h_0$ is a constant determined by a preliminary fit. To test the reliability of inferring the vortex energy based solely on vortex positions, we have also numerically generated random (unclustered) ensembles of $N_{\pm} = 8$ vortices, equal to the mean number of vortices detected within the 89\% detection region. We then calculate the energy of the configuration assuming that all vortices on the top left (bottom right) half of the ellipse have positive (negative) circulation.  This energy is indicated in the insets of Figs.~\ref{SuppEnergyDecay}, {A and B} as horizontal shaded regions, representing a 95\% confidence interval corresponding to the 10 samples at each hold time ($\pm1.96\sigma/\sqrt{10}$).  For the times the experimentally estimated vortex energy is larger than this value, we are confident that the vortices remain clustered despite the loss of vortex energy.

We observe a sharp reduction in the cluster energy decay time for condensate fractions below 65\% (Fig.~\ref{SuppEnergyDecay}{A}), which along with the nearest-neighbor analysis,~Fig.~4, confirms thermal friction as a primary source of dissipation. We also find that increasing the density variation leads to apparent increased energy damping, shown in Fig.~\ref{SuppEnergyDecay}{B}, in contrast to the nearest-neighbor behavior in Fig.~\ref{SuppNNDecay}. While suggesting some increased energy loss due to the increased density variation, we note that the nearest-neighbor behavior indicates relatively tight vortex clusters are maintained. We note that the the decay fit for the largest $\{160,115\}~\mu\textrm{m}$ trap (Fig.~\ref{SuppEnergyDecay}{B} inset) tends towards an energy value above the uncorrelated estimate. In conjunction with the vortex position histogram shown in Fig.~\ref{SuppEnergyDecay}{A}, we speculate that this may indicate the emergence of a monopole state, consisting of a central like-circulation cluster surrounded by opposite circulation vortices and possessing net angular momentum. In a weakly elliptical trap this state will have a comparable entropy to the (maximal entropy) dipole configuration~(\emph{21,48}).

\clearpage \newpage

\renewcommand{\thefigure}{S\arabic{figure}}
\setcounter{figure}{0}  

\newpage
\begin{figure*}
		\centering\includegraphics[width=.6\textwidth]{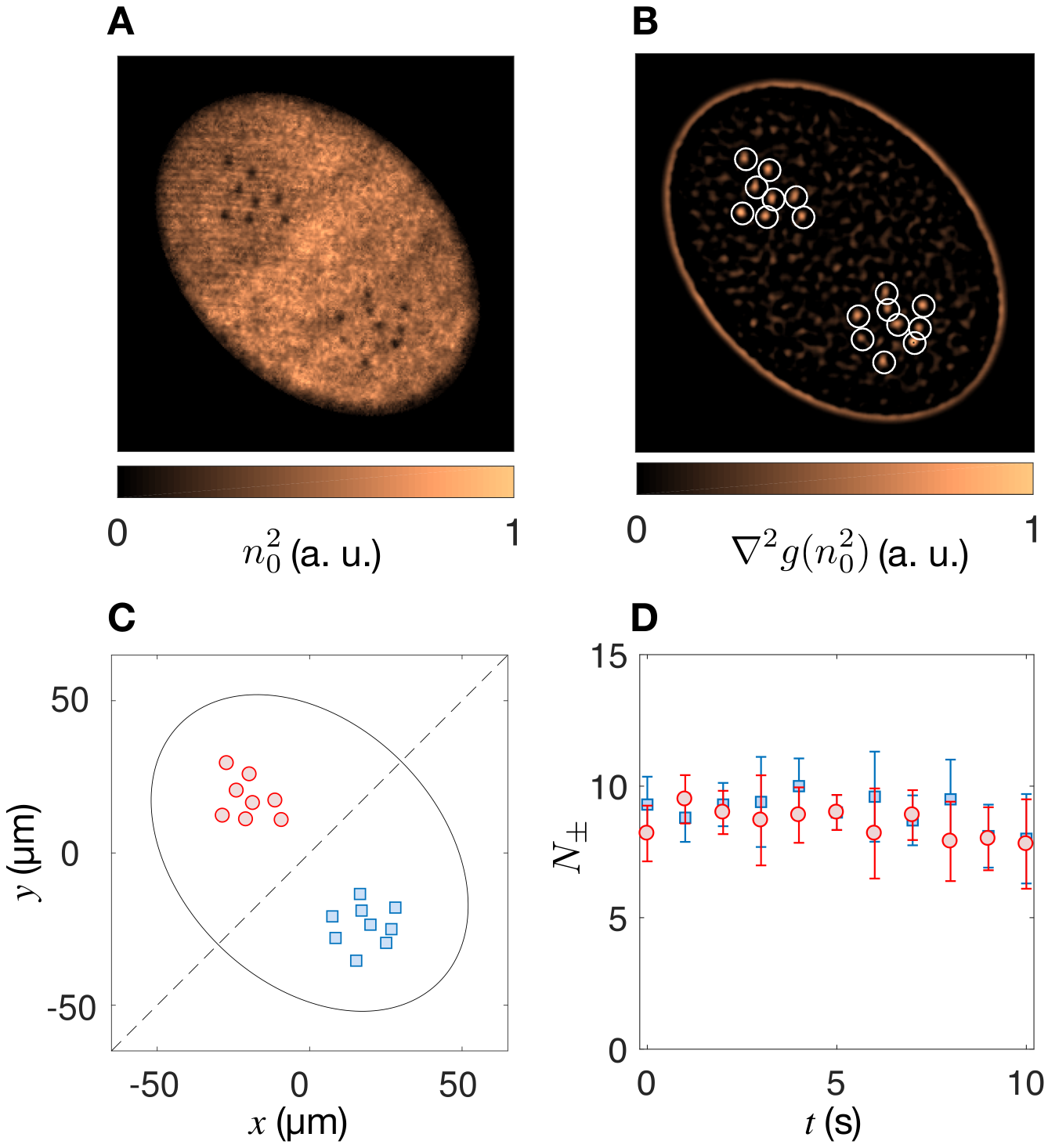}
		\caption{\textbf{Vortex fitting and classification.} \textbf{A,} Background-subtracted Faraday image of the experimental BEC column density. \textbf{B,} The Gaussian blob algorithm takes the Laplacian of the Gaussian-filtered image to locate the vortex cores. \textbf{C,} Vortex circulations are then assigned across the minor axis, with positive vortices indicated by red circles and negative vortices indicated by blue squares. \textbf{D,} The distribution of positive and negative assigned vortices as a function of time for the data sets corresponding to Fig.~3A. Nearly equal numbers of positive and negative vortices are obtained throughout the hold times. The error bars indicate the standard deviation of the data.}
		\label{SuppDetection}

\end{figure*}

\begin{figure*}
		\centering\includegraphics[width=.5\textwidth]{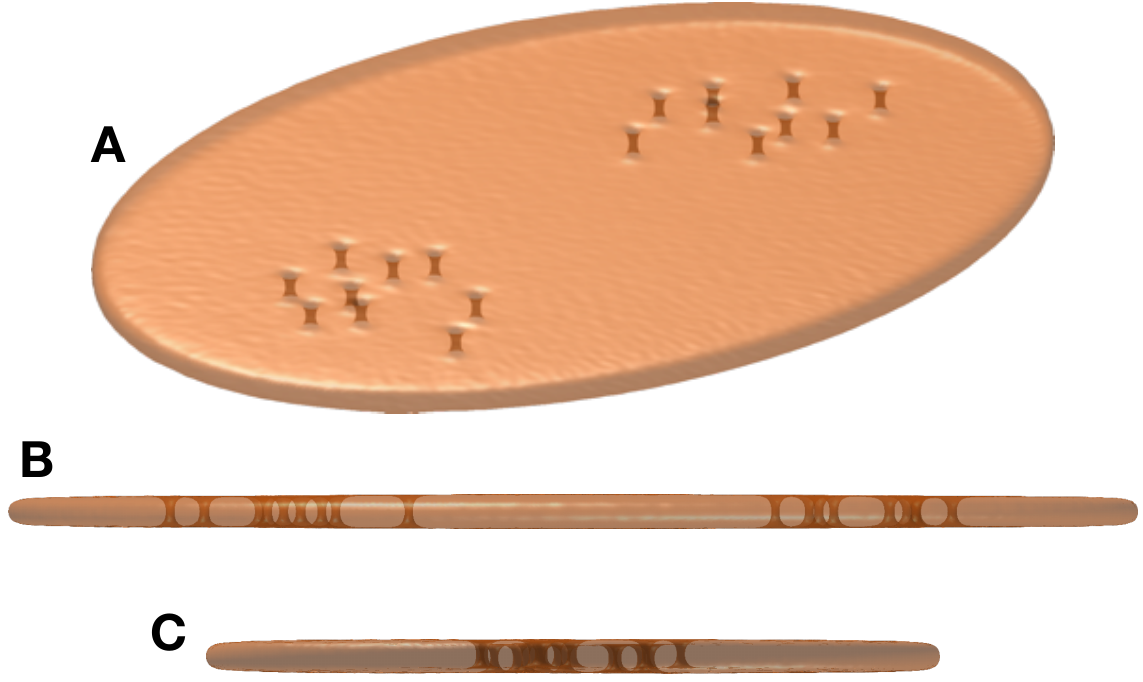}
		\caption{\textbf{3D GPE simulation data.} Density isosurfaces (approximately $25\%$ of peak density) are shown shortly after the paddle stir: \textbf{A,} angle view, \textbf{B,} view looking down the minor axis, \textbf{C,} view looking down the major axis. The vortices are clearly rectilinear; no vortex bending is visible.}
		\label{Supp3DGPE}

\end{figure*}

\begin{figure*}
		\centering\includegraphics[width = \textwidth]{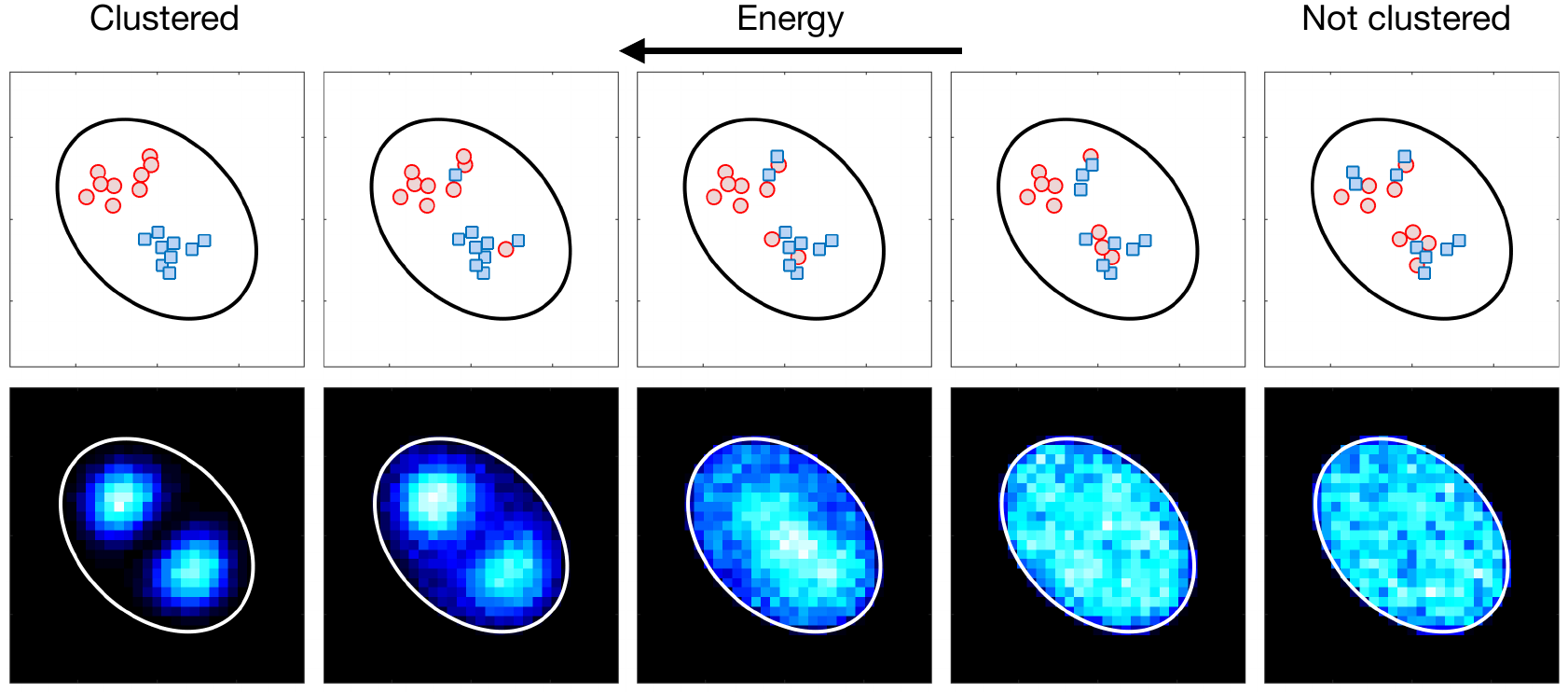}
		\caption{\textbf{Point-vortex dynamics.} Time averaged vortex densities produced from point vortex dynamics (bottom row), using a sample from the paddle experiment for the initial positions (top row). The leftmost example assumes the clusters are all of the same sign, whereas the rightmost assumes the charges are random. Intermediate cases show the effect of selecting 1, 2, or 3 vortices from each cluster at random and swapping their signs to reduce the cluster net charge and lower the energy. Note that the initial positions are identical for all initial conditions; only the vortex signs are different.}
		\label{fig:timeAverageDensity}
\end{figure*}

\begin{figure*}
	\centering\includegraphics[width = \textwidth]{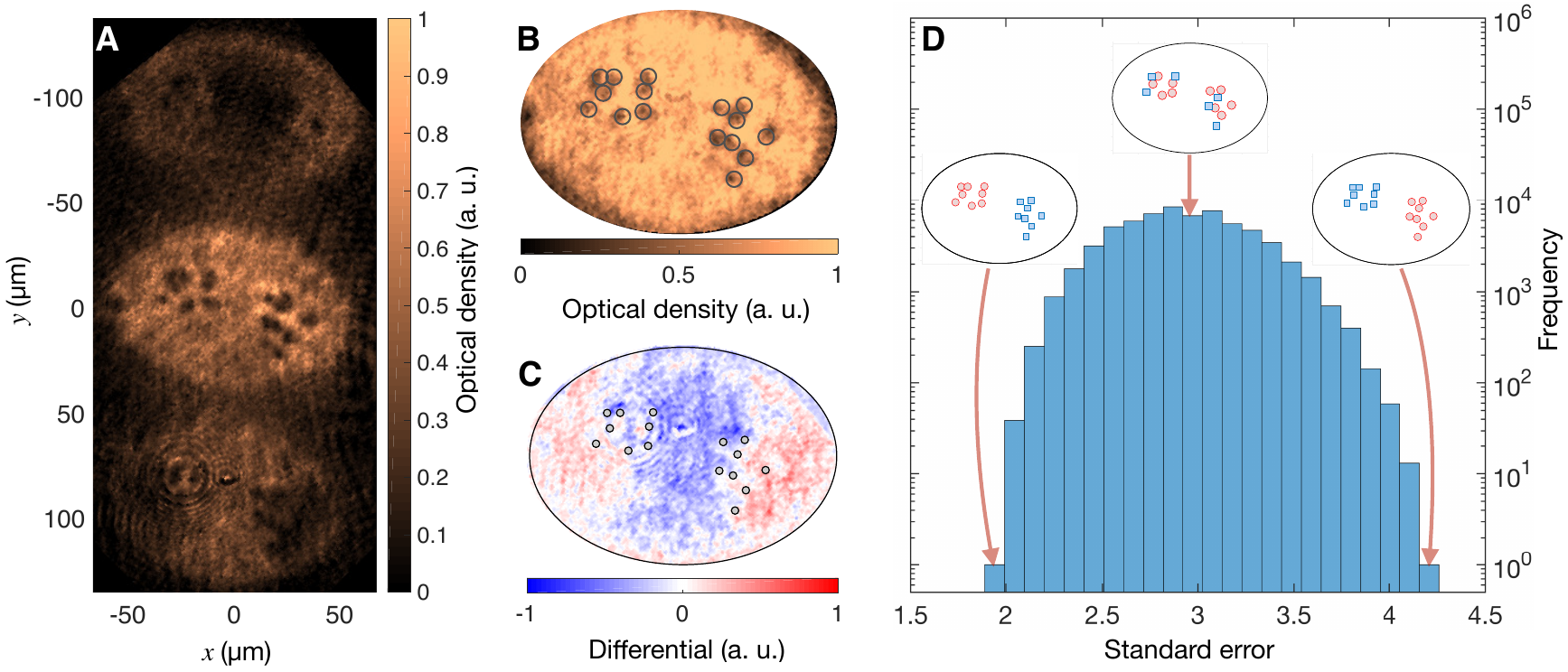}
		\caption{\textbf{Bragg-scattering vortex sign detection.} \textbf{A,} Absorption imaging of the condensate after cluster injection and Bragg scattering, followed by a 10~ms TOF, showing the Bragg-scattered components and central unscattered cloud. \textbf{B,} Vortex detection on the sum of unscattered and scattered components; the corresponding differential signal of the scattered components is shown in~\textbf{C}. \textbf{D,} Vortex signs are permuted through the $2^{16}$ charge configurations for the 16 vortices (examples shown in insets), and the standard error between the experimental,~\textbf{C}, and a synthesized differential signal is calculated (see text); the histogram demonstrates that a single configuration minimizes the error, containing single sign clusters as in Fig.~2C. Swapping the vortex signs results in the standard error being maximized.}
		\label{fig:Bragg}
\end{figure*}

\begin{figure*}
		\centering\includegraphics[width=.7\textwidth]{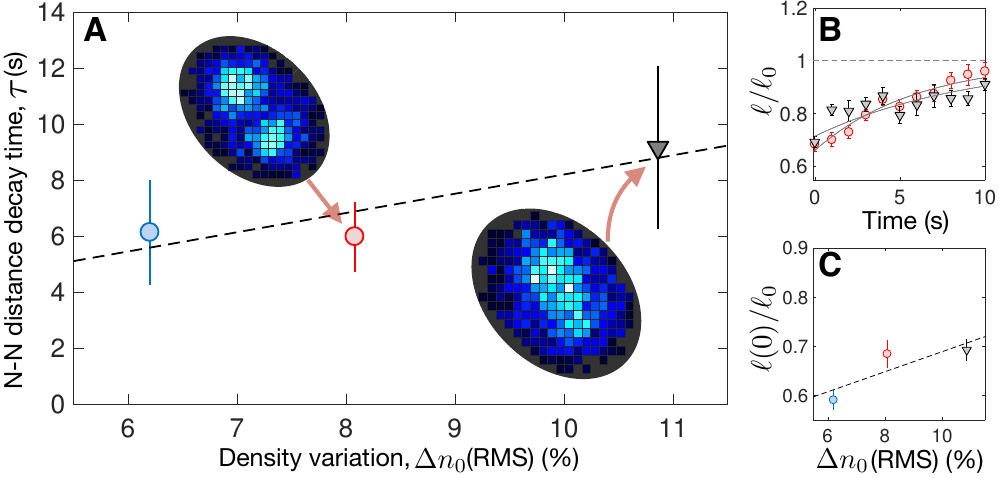}
		\caption{\textbf{Cluster decay rates with increased density variation.} \textbf{A,} Increasing the ellipse size results in the residual harmonic confinement becoming more significant, leading to increased density variation, while maintaining high condensate fractions. Nearest-neighbor distance decay times are determined by exponential fits (see text), \textbf{B,} with the $\{2a,2b\} = \{125,85\}~\mu$m trap (blue circles),  $\{140,100\}~\mu$m trap (red circles), and largest $\{160,115\}~\mu$m trap (grey triangles) shown. Insets show time-averaged vortex density histograms accumulated over a 10 second hold for larger cases, as in Fig.~4; see Fig.~\ref{SuppInitHist} for the histograms immediately following the stir. \textbf{C,} The initial nearest-neighbor distance $\ell(0)$ varies over a smaller range when compared with Fig.~4. Dashed lines indicate linear fits to the data.}
		\label{SuppNNDecay}
\end{figure*}

\begin{figure*}
		\centering\includegraphics[width=1\textwidth]{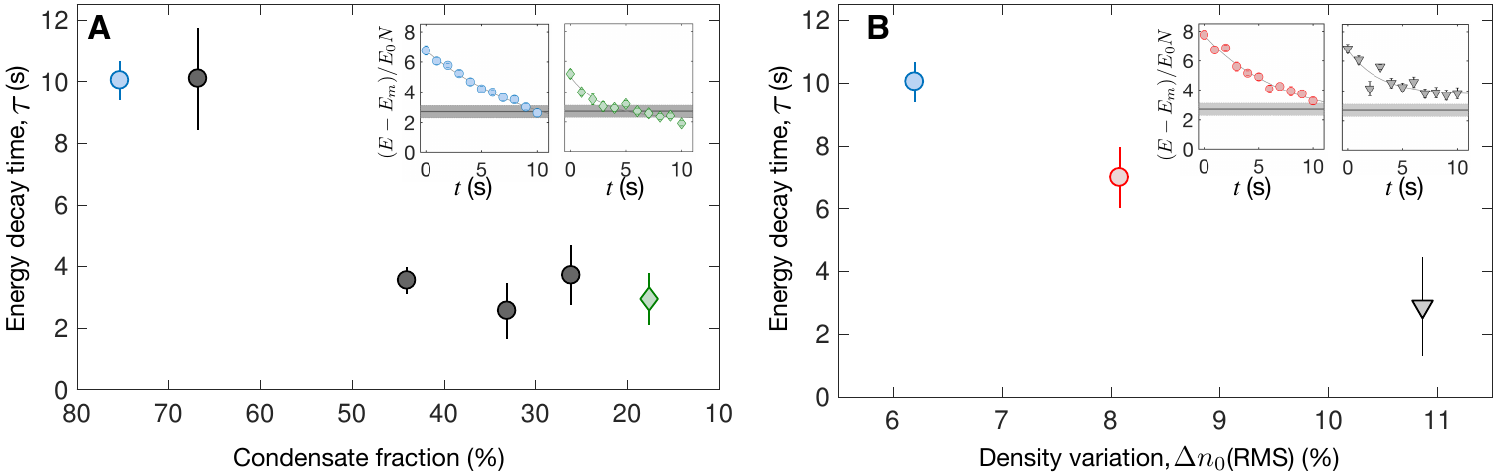}
		\caption{\textbf{Energy damping rates for varying BEC fraction and non-uniform density.} \textbf{A,} Energy decay times for varying BEC fraction, displaying a decrease in damping time with increased thermal fraction, consistent with nearest-neighbor and histogram analysis. (Insets) Energy versus hold time for the largest (blue circles) and smallest (green diamonds) condensate fractions.  \textbf{B,} Energy decay times with increasing non-uniform density, where the leftmost point corresponds to the $\{2a,2b\}= \{120,85\}~\mu$m trap. (Insets) The decay of the vortex energy for the intermediate (red circles) and largest (black triangles) traps.  The shaded region indicates the upper bound of the vortex configuration energy if our circulation allocation algorithm was applied to a random vortex ensemble (see text). The lifetime is determined by fits to offset exponential decays,  shown with dash-dot lines.}
		\label{SuppEnergyDecay}
\end{figure*}

\begin{figure*}
		\centering\includegraphics[width=\textwidth]{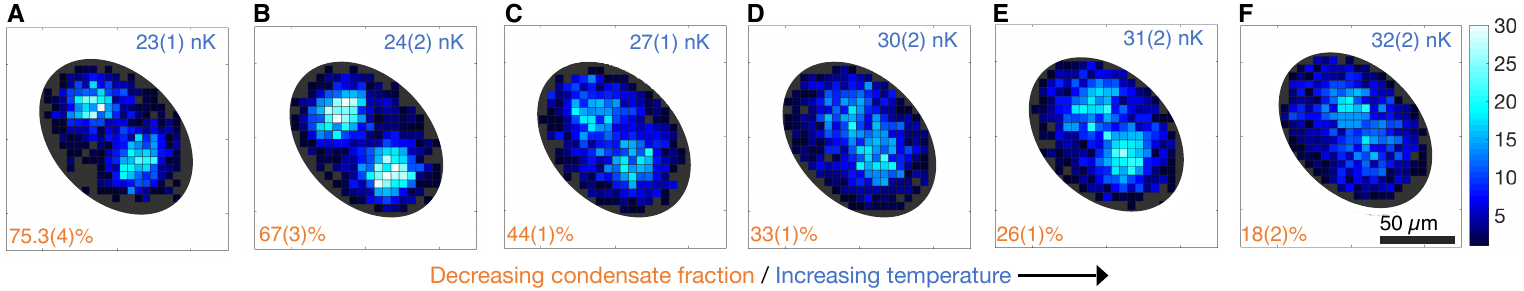}
		\caption{\textbf{Time-averaged vortex position histograms as a function of condensate fraction.} \textbf{A -- F,} Vortex position histograms corresponding to the full condensate fraction and temperature range in the $\{2a,2b\} = \{120,85\}~\mu$m trap considered in Fig.~4 of the main text.  The initial condensate fraction is indicated in the bottom left, and the temperature in the top right of each subfigure.}
		\label{SuppHist}
		\end{figure*}

\begin{figure*}
		\centering\includegraphics[width=.8\textwidth]{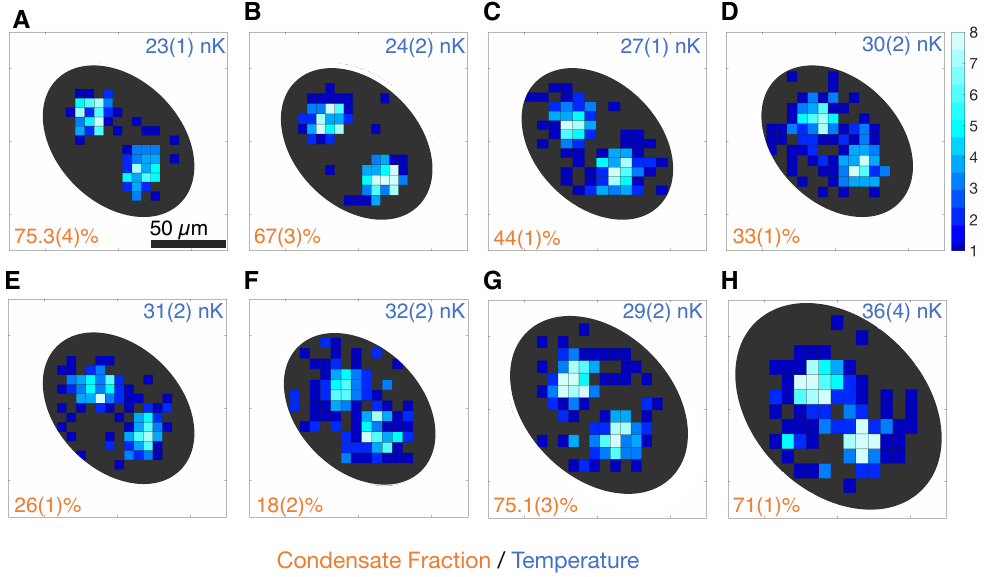}
		\caption{\textbf{Initial position vortex histograms immediately after the stir.} \textbf{A -- F,} Vortex position histograms corresponding to the full condensate fraction and temperature range in the $\{2a,2b\}= \{120,85\}~\mu$m trap considered in Fig.~4 of the main text. \textbf{G,~H} Vortex position histograms for the $\{140,100\}~\mu$m trap, and the $\{160,115\}~\mu$m trap.  The initial condensate fraction is indicated in the bottom left, and the temperature in the top right of each subfigure.}
		\label{SuppInitHist}

\end{figure*}

\begin{figure*}
		\centering\includegraphics[width=.8\textwidth]{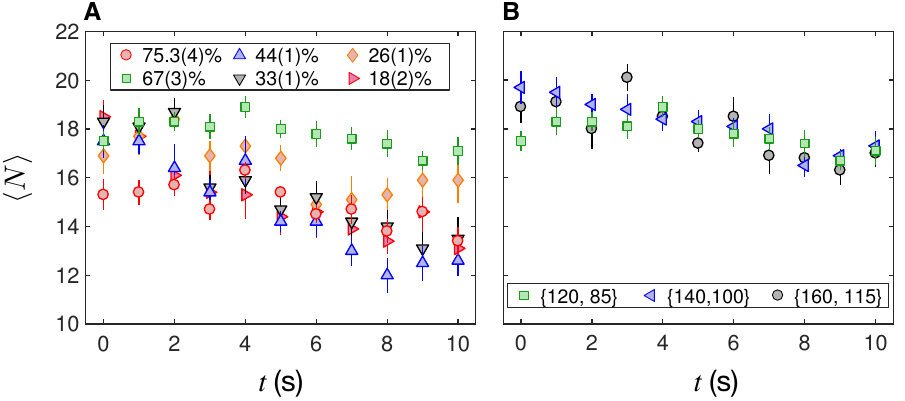}
			\caption{\textbf{Mean vortex number vs. hold time for the paddle stir.} \textbf{A}, Vortex numbers as a function of time for different condensate fractions: 75.3(4)\% (red circles), 67(3)\% (green squares), 44(1)\% (blue upward facing triangles), 33(1)\% (black downward facing triangles), 26(1)\% (orange diamonds), 18(2)\% (red right facing triangles). \textbf{B}, Vortex numbers as a function of time for the larger traps, with larger condensate density variations: $\{2a,2b\} = \{120,85\}~\mu$m trap and 67(3)\% fraction (green squares), $\{140,100\}~\mu$m trap (blue left facing triangles), and $\{160,115\}~\mu$m trap (black circles).}
		\label{SuppVortexNumbers}
\end{figure*}

\end{document}